\documentclass[aps,prl,showpacs,preprintnumbers,amssymb,superscriptaddress,twocolumn]{revtex4-1}
\usepackage[dvipdfmx]{graphicx}
\usepackage{bm}
\usepackage{amsmath}
\usepackage{ascmac}
\usepackage{setspace}
\usepackage{amssymb}
\usepackage{latexsym}
\usepackage{mathtools}
\usepackage{color}

\begin{document}

\title{Flemish Strings of Magnetic Solitons and a Non-Thermal Fixed Point in a One-Dimensional Antiferromagnetic Spin-1 Bose Gas} 

\author{Kazuya Fujimoto}
\affiliation{Department of Physics, University of Tokyo, 7-3-1 Hongo, Bunkyo-ku, Tokyo 113-0033, Japan}

\author{Ryusuke Hamazaki}
\affiliation{Department of Physics, University of Tokyo, 7-3-1 Hongo, Bunkyo-ku, Tokyo 113-0033, Japan}

\author{Masahito Ueda}
\affiliation{Department of Physics, University of Tokyo, 7-3-1 Hongo, Bunkyo-ku, Tokyo 113-0033, Japan}
\affiliation{RIKEN Center for Emergent Matter Science (CEMS), Wako, Saitama 351-0198, Japan}

\date{\today}

\begin{abstract}
Thermalization in a quenched one-dimensional antiferromagnetic spin-1 Bose gas is shown to proceed via a non-thermal fixed point through annihilation of Flemish-string bound states of magnetic solitons. A possible experimental situation is discussed.
\end{abstract}


\maketitle

{\it Introduction.-}
Ultracold atomic gases offer an ideal playground for studying universal nonequilibrium dynamics due to their high controllability and isolation from the environment \cite{Pol11,Eisert15,Altman}. Indeed, fundamental aspects of universal thermalization dynamics near critical points at equilibrium phase transitions have been studied in terms of the Kibble-Zurek mechanism \cite{KZ1,KZ2,KZ3,KZ4,KZ5,KZ6} and dynamical critical phenomena \cite{DPT1}. Even far from critical points, isolated systems are found to exhibit universal thermalization in the decay of turbulence \cite{NTF1,NTF2} and coarsening dynamics \cite{CDinBEC0,CDinBEC1,CDinBEC2,CDinBEC3,CDinBEC4,CDinBEC5,CDinBEC6,CDinBEC6_1,CDinBEC7,CDinBEC8,CDinBEC9,CDinBEC9_1,CDinBEC10,CDinBEC11,CDinBEC12,CDinBEC13,CDinBEC14,CDinBEC15,CDinBEC16}. However, unlike the dynamics near critical points \cite{DCP1,DCP2}, a unified framework for understanding the universal behavior is yet to be established.

\begin{figure}[b]
\begin{center}
\includegraphics[keepaspectratio, width=8.8cm,clip]{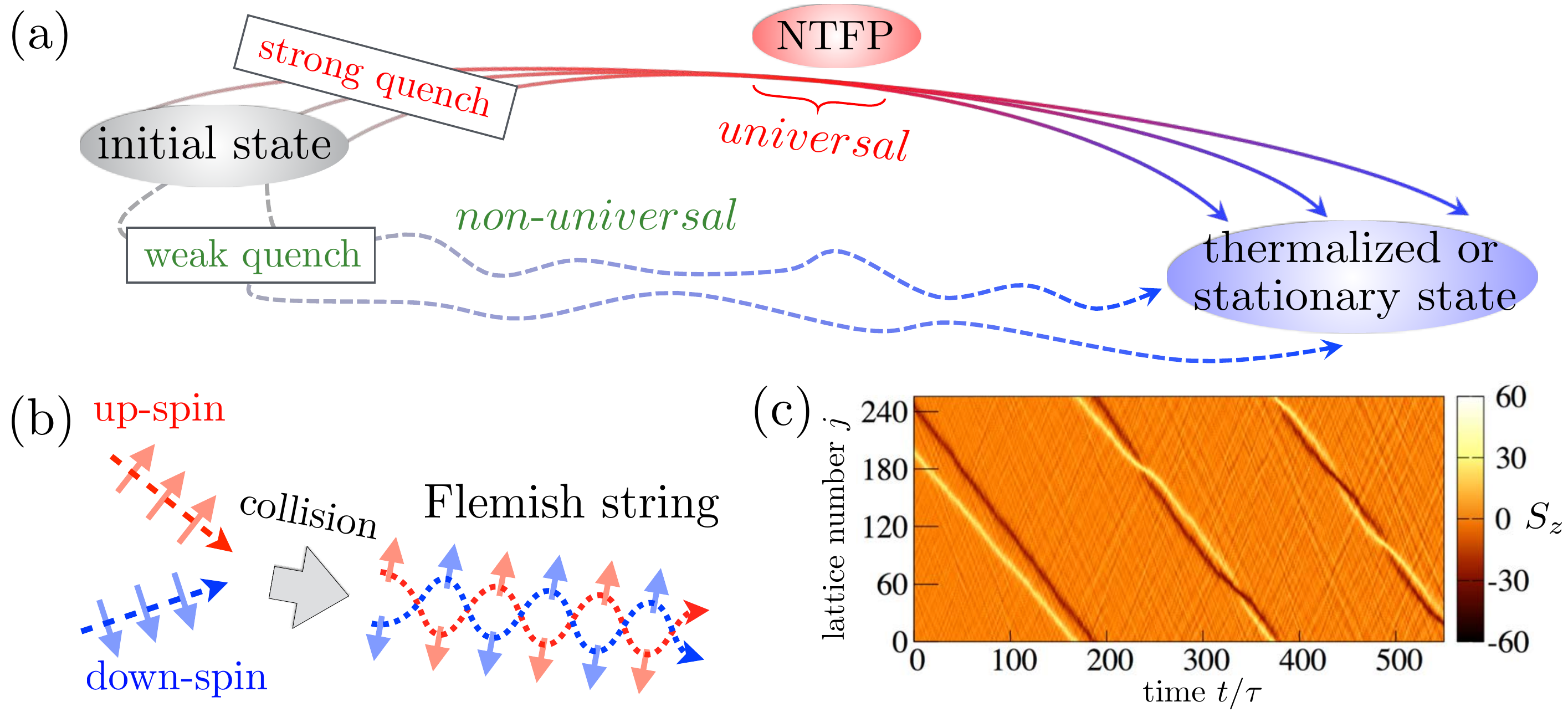}
\caption{(a) Schematic illustration of a NTFP. An isolated strongly quenched quantum system undergoes thermalization via a NTFP, where the universal scaling relation \eqref{NTFP} emerges en route to a thermalized or stationary state. However, a weakly quenched system may undergo non-universal thermalization. (b) Schematic illustration of a Flemish string. A collision between two magnetic solitons with opposite signs of magnetization creates a Flemish string, which has a twisted magnetic structure. The solid (dashed) arrows show the direction of magnetization (trajectories of magnetization). (c) Spatio-temporal distribution of $\hat{S}_{z,j}$ showing formation of a Flemish string, which is obtained from a mean-field numerical calculation of Eq.~\eqref{spinor_BH} subject to the periodic boundary condition. 
\label{intro} }
\end{center}
\end{figure}

Recently, a non-thermal fixed point (NTFP) \cite{NTF1,NTF2,NTF3,NTF4} has attracted growing interest as a universal thermalization scenario in isolated quantum systems. It is expected to provide a unified framework of different nonequilibrium phenomena in diverse systems ranging from cosmology to cold atoms~\cite{NTF3,NTF4}. As illustrated in Fig.~\ref{intro}(a), a system with a strongly quenched initial state undergoes relaxation via a NTFP at which the scale-invariant thermalization dynamics emerges transiently. The NTFP is characterized by a universal scaling of an equal-time correlation function $C({\bm x},t)$ as $C({\bm x},t) = s^{-\gamma} C({\bm x}s^{\beta},ts)$ with an arbitrary parameter $s$ and scaling exponents $\beta$ and $\gamma$. Setting $s=t^{-1}$, we obtain the dynamical scaling:
\begin{eqnarray}
C({\bm x},t) = t^{\gamma} f({\bm x}t^{-\beta}), \label{NTFP}
\end{eqnarray}
where $f({\bm y}) = C({\bm y},1)$ is a universal function. After passing through the NTFP, the system relaxes toward a thermalized or stationary state. 
This scenario has been studied theoretically in two- and three-dimensional Bose gases \cite{NTF_GP1,NTF_GP2,NTF_GP3,NTF_GP4,NTF_GP5,NTF_GP6,NTF_GP7} and a relativistic $O(n)$ model \cite{NTF_O1,NTF_O2,NTF_O3} in the context of wave turbulence \cite{WT1,WT2,WT3} and coarsening dynamics  \cite{coar1,coar2}. Here weakly interacting waves or topological objects such as vortices and domain walls play essential roles in thermalization processes. Very recently, the universal scaling in Eq.~\eqref{NTFP} in momentum space was observed in one-dimensional (1D) ultracold atomic gases \cite{NTF_exp1,NTF_exp2}. However, it remains to be fully understood under what conditions a system shows universal behavior especially in an experimentally controllable manner. For example, a weak quench in Fig.~\ref{intro}(a) generates a small number of elementary excitations and may become thermalized without passing through a NTFP.

\begin{figure}[t]
\begin{center}
\includegraphics[keepaspectratio, width=8.8cm,clip]{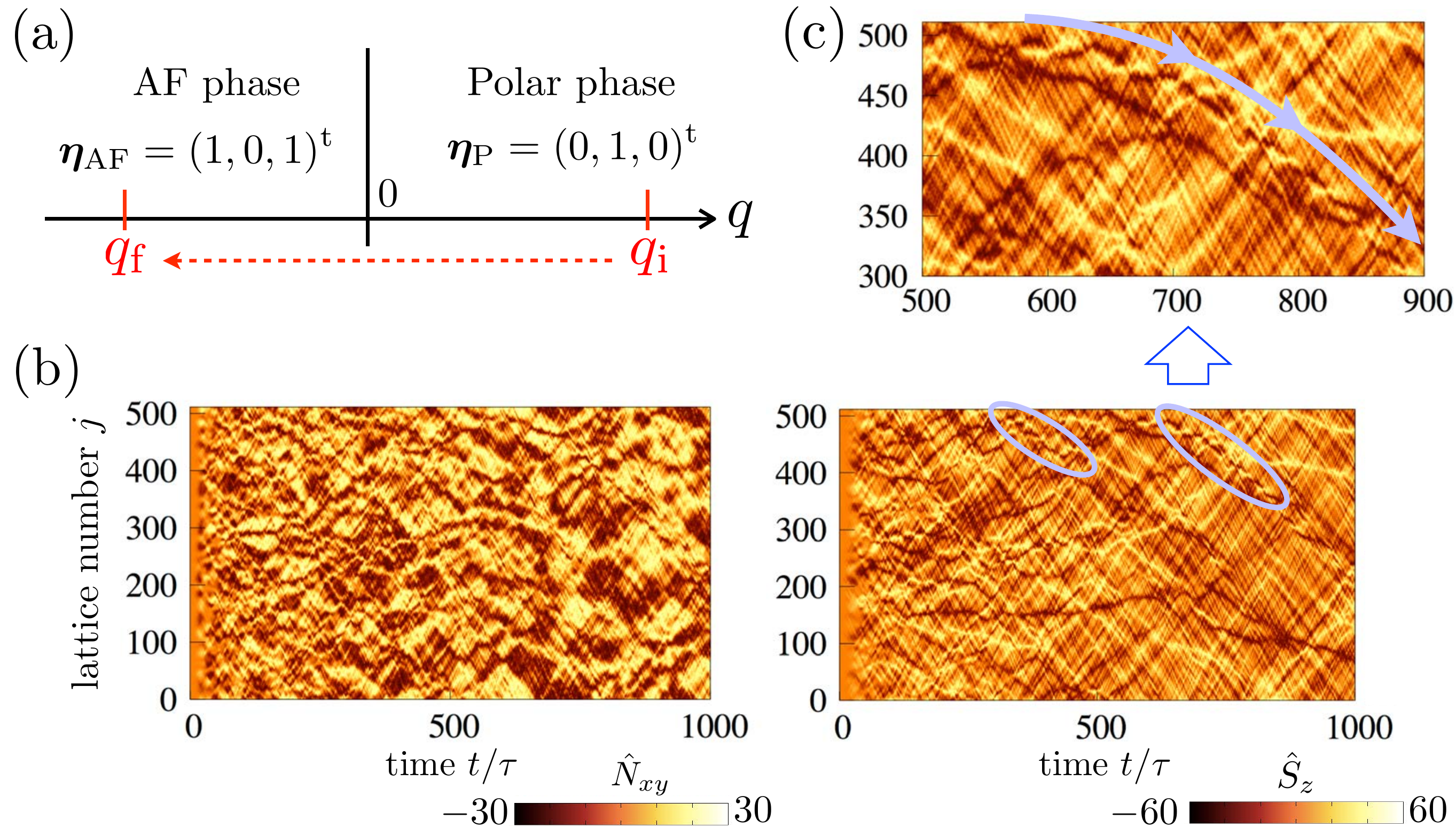}
\caption{ (a) Quench protocol. A spin-1 antiferromagnetic (AF) Bose gas with a quadratic Zeeman term has two mean-field ground states: AF and polar, where $\bm{\eta}_{\rm AF}$ and $\bm{\eta}_{\rm P}$ are the corresponding spinor wavefunctions. We suddenly quench the strength $q$ of the quadratic Zeeman term from $q_{\rm i}$ to $q_{\rm f}$. (b) Spatio-temporal distributions of nematicity $\hat{N}_{xy,j}$ (left) and spin $\hat{S}_{z,j}$ (right) calculated from a single trajectory in the TWA with $U_2/U_0=0.050$, $\Gamma = 2$, and $\tau=4\hbar/J$. The left panel shows the growth of nematic domains. The bright and dark lines in the right panel show positively and negatively charged magnetic solitons, respectively. The number of solitons decreases in time (see also Fig. 4(c)) as Flemish strings enclosed by solid circles are annihilated. (c) Enlarged figure of $\hat{S}_{z,j}$ in (b). The arrow shows the direction of motion of a Flemish string. \label{distribution} }
\end{center}
\end{figure}

In this Letter, we theoretically study quench dynamics in a 1D spin-1 Bose gas with an antiferromagnetic (AF) interaction, unveiling universal thermalization dynamics through a NTFP caused by exotic bound states of twisted magnetic solitons, which we refer to as Flemish strings (see Figs.~\ref{intro}(b) and (c)). Since the spin-exchange interaction is AF, the relaxation dynamics after the quench in Fig.~\ref{distribution}(a) is characterized by nematicity \cite{Nematic1,Nematic2,Nematic3,Nematic4}. We find that a nematic correlation function transiently exhibits the dynamical scaling \eqref{NTFP} with $\beta \simeq 0.32$ and $\gamma \simeq 0.11$. The underlying physics of this universal thermalization through a NTFP is highly nontrivial cooperative soliton dynamics, where two magnetic solitons, which are stable individually \cite{magsoli1,magsoli2,magsoli3,magsoli4}, undergo pair-annhilation by forming Flemish strings. While there are a number of studies for quench dynamics of an AF spinor Bose gas \cite{CDinBEC9_1,quench0,quench1,quench2,quench3,quench4,quench5,quench6,quench7,quench8,quench9,quench10}, universal thermalization through Flemish strings has never been reported. 

Our results reported here offer the first clear evidence of the soliton-induced NTFP. In Ref.~\cite{NTF_pre}, a NTFP in a 1D Bose gas is theoretically discussed, but the dynamical scaling \eqref{NTFP}, which is regarded as a hallmark of a NTFP \cite{NTF3}, is not confirmed. In Ref.~\cite{NTF_exp2}, a relation between the observed universal thermalization dynamics and a random grey soliton model in a 1D repulsive Bose gas \cite{NTF_pre} is discussed. However, the observed exponent $\beta$ deviates by 50\% from the exponent obtained by assuming the theoretical model, and the other key exponent $\alpha = \beta+\gamma$ is not explained from the perspective of solitons. Thus, the relation between the NTFP and solitons has remained unclear. 

Furthermore, we address an unsolved issue raised in Refs.~\cite{AF_quench1,AF_quench2}, where thermalization dynamics after the same quench as ours in Fig.~\ref{distribution}(a) was experimentally studied in trapped 1D AF Bose gases of $^{23}{\rm Na}$. The experiments investigate the observed dynamics in terms of coarsening \cite{coar1,coar2}, but universal aspects of thermalization have remained an unsolved issue. We argue that the universal thermalization dynamics characterized by the NTFP did not emerge in the experiments of Refs.~\cite{AF_quench1,AF_quench2} because the quench was so weak that only a few solitons were created, and show that the NTFP due to Flemish strings should emerge for stronger quench. 

{\it Universal~thermalization~dynamics~in~a~uniform~system.-}
We consider a translation-invariant spin-1 Bose-Hubbard model \cite{SBH} subject to the periodic boundary condition with the lattice constant $a$ and the number of lattices $M$. We denote the annihilation (creation) operator of spin-1 bosons with magnetic quantum number $m$ at site $j$ as $\hat{b}_{m,j}$ ($\hat{b}^{\dagger}_{m,j}$). Then, the Hamiltonian is given by
\begin{eqnarray}
\hat{H}_{\rm BH} = &-&J \sum_{m,j} \Bigl(\hat{b}_{m,j+1}^{\dagger} \hat{b}_{m,j} + {\rm h.c.} \Bigl) + q \sum _{m,j} m^2\hat{b}^{\dagger}_{m,j} \hat{b}_{m,j} \nonumber \\
&+& \frac{U_0}{2} \sum_{j} \hat{\rho}_j(\hat{\rho}_j-1) + \frac{U_2}{2} \sum_j \Bigl( \hat{\bm{S}}_j^2 - 2 \hat{\rho}_j \Bigl) \label{spinor_BH}
\end{eqnarray}
with the hopping parameter $J$, the quadratic Zeeman coefficient $q$, the density interaction coefficient $U_0$, and the spin interaction coefficient $U_2$. Here, $\hat{\rho}_j \coloneqq \sum_{m}  \hat{b}_{m,j}^{\dagger} \hat{b}_{m,j}$ is the total particle-number operator and $\hat{S}_{\alpha,j} \coloneqq \sum_{m,n}  \hat{b}_{m,j}^{\dagger} ({s_{\alpha}})_{mn} \hat{b}_{n,j}~(\alpha=x,y,z)$ with the spin-1 spin matrices $(s_{\alpha})_{mn}$ is the spin-vector operator. 

We focus on the AF interaction ($U_2>0$), so that the mean-field ground-state phase is either polar ($q>0$) or AF ($q<0$) (see Fig.~\ref{distribution}(a)). Since both phases are non-magnetic, the spinor order parameter is a second-rank tensor of the spin matrices \cite{Nematic1,Nematic2,Nematic3} described by the nematic tensor $\mathcal{N}_{\mu \nu,j} \coloneqq \langle \hat{N}_{\mu \nu,j} \rangle$ with $\hat{N}_{\mu \nu,j} \coloneqq \frac{1}{2} \sum_{l,m,n}\hat{b}_{m,j}^{\dagger}  \Bigl[(s_{\mu})_{mn}(s_{\nu})_{nl}  + (\mu \leftrightarrow \nu)\Bigl] \hat{b}_{l,j}$. Here, $\langle \cdots \rangle$ means the average over the state vector $| \psi(t) \rangle $.
In the AF phase, the nematic tensor becomes
\begin{eqnarray}
  \mathcal{N}^{\rm AF}_{\mu \nu} = \frac{N}{2M}\left(
    \begin{array}{ccc}
      1-{\rm cos(2 \alpha)} & -{\rm sin(2 \alpha)} & 0 \\
      -{\rm sin(2 \alpha)} & 1-{\rm cos(2 \alpha)} & 0 \\
      0 & 0 & 2
    \end{array}
  \right),  \label{AF_order}
  \end{eqnarray}
where $\alpha$ is the azimuthal angle in spin space and $N$ is the number of condensed bosons. When we quench $q$ from the polar phase to the AF phase as shown in Fig.~\ref{distribution}(a), the system gets highly excited and starts relaxation.

We employ the truncated Wigner approximation (TWA) \cite{TWA0,TWA1,TWA2} to numerically study the thermalization dynamics of a quenched AF Bose gas with $J/U_{0}=40$, $M=512$, and $N=2 \times 10 ^{4}$, with which the system is in a deep superfluid regime where the TWA should work well \cite{TWA1,TWA2}. The initial state is chosen to be the Bogoliubov vacuum for the polar phase \cite{CDinBEC14}, and we suddenly quench $q$ from $q_{\rm i}=NU_{2}/M$ to $q_{\rm f} = (1-\Gamma)NU_{2}/M$ where $\Gamma$ controls the strength of the quench. 

Figure~\ref{distribution}(b) shows the spatio-temporal distributions of nematicity $\hat{N}_{xy,j}$ (left) and magnetization $\hat{S}_{z,j}$ (right) obtained from a single trajectory. The typical domain size of $\hat{N}_{xy,j}$ grows in time, while the number of locally magnetized domains, which are magnetic solitons \cite{magsoli1,magsoli2,magsoli3,magsoli4}, decreases in time (see Fig.~\ref{solitons}(c)). As shown in Fig.~\ref{distribution}(c), magnetic solitons can form a Flemish string in which two magnetic solitons form a virtual bound state and evolve in time in a twisted manner. The formation of a Flemish string is essential for thermalization as discussed later. 

\begin{figure}[t]
\begin{center}
\includegraphics[keepaspectratio, width=8.8cm,clip]{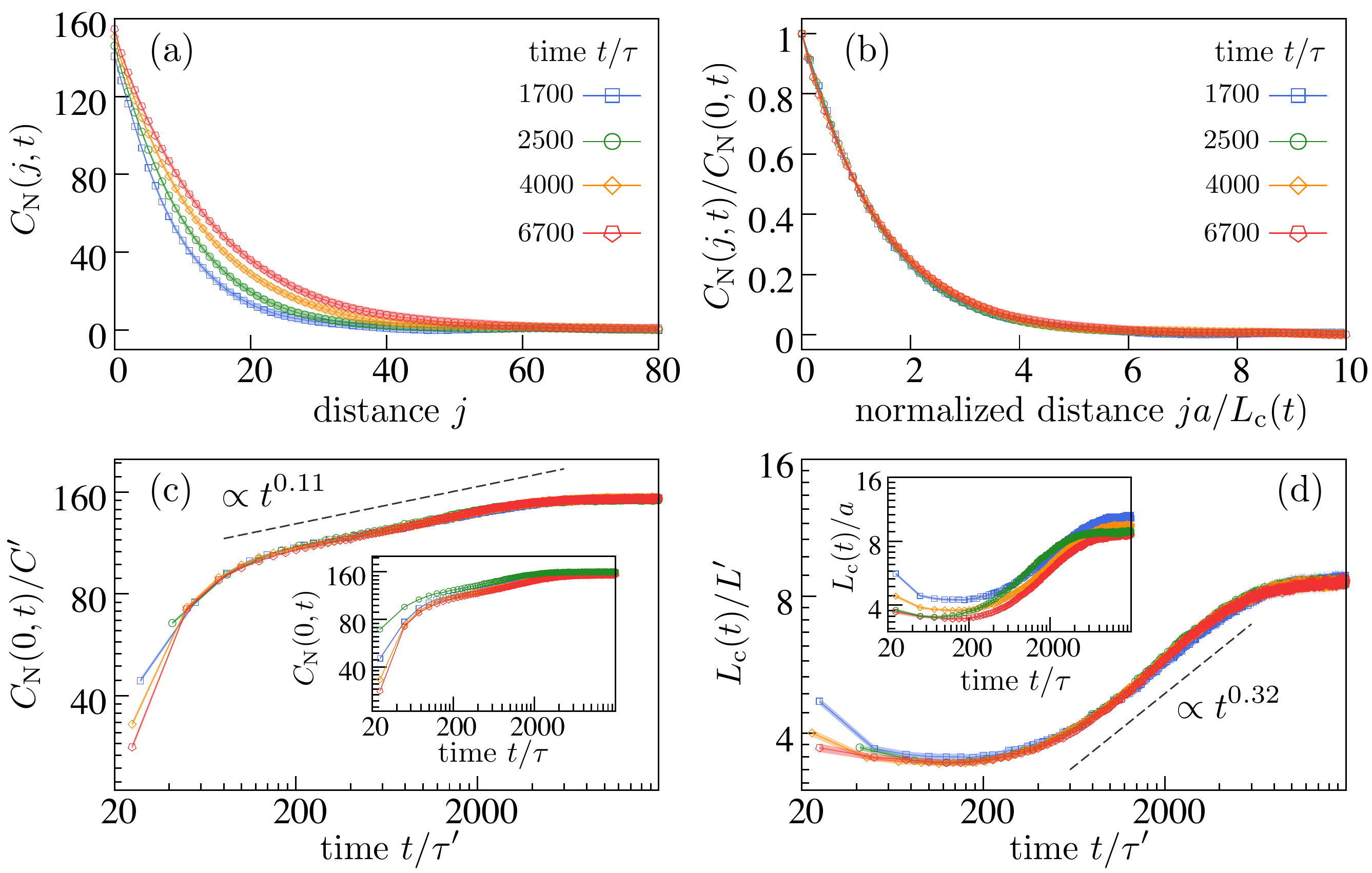}
\caption{(a) Time evolution of a correlation function $C_\mathrm{N}(j,t)$ for $U_2/U_0=0.05$ and $\Gamma=2$, and (b) the same quantity with the abscissa and the ordinate normalized by $L_\mathrm{c}(t)/a$ and $C_\mathrm{N}(0,t)$, respectively. In (b), all curves collapse to a single universal curve consistent with the dynamical scaling described by Eq.~\eqref{NTFP}. (c,d) Time evolution of (c) $C_{\rm N}(0,t)$ and (d) $L_{\rm c}(t)$ for four sets of spin interaction and quench strength parameters: ($U_2/U_0$, $\Gamma$)= (0.075, 2)[green circle], (0.05, 2)[blue square], (0.05, 2.2)[yellow diamond], and (0.05, 2.4)[red pentagon]. The insets show the same results for the unnormalized axes. For comparison, all axes in the main panels of (c,d) are multiplied by different constants $L',C',\tau'$ for each run. These results show that the exponents $\beta\simeq 0.32$ and $\gamma \simeq 0.11$ are universal and independent of system's parameters. The power laws transiently appear, and finally disappear when the system is close to a stationary state, which is consistent with Fig.~\ref{intro}(a). Color bands show $3\sigma$ error bars in the TWA calculation. \label{correlation_length}}
\end{center}
\end{figure}

To investigate the universal scaling in Eq.~\eqref{NTFP}, we calculate the nematic correlation function defined by
\begin{eqnarray}
C_{{\rm N}}(j,t) \coloneqq \frac{1}{M}\sum_{k=0}^{M-1} \langle \hat{N}_{xy,j+k}\hat{N}_{xy,k} \rangle(t).
\label{Ncorrelation_function}
\end{eqnarray}
Here, we consider the $xy$-component of $\hat{N}_{\mu \nu,j}$ because Eq.~\eqref{AF_order} shows that the order parameter of the AF phase can be characterized  by $\alpha$ alone. The time evolution of $C_{{\rm N}}(j,t)$ shows that the correlation length $L_{\rm c}(t)$ monotonically increases as shown in Fig.~\ref{correlation_length} (a). Here, $L_{\rm c}(t)$ is determined from $C_{\rm N}(j=L_c(t)/a,t)=0.5C_{\rm N}(j=0,t)$. When the abscissa and the ordinate are normalized by $L_{\rm c}(t)/a$ and $C_{{\rm N}}(0,t)$, respectively, the correlation functions at different times collapse into a single universal curve as shown in Fig.~\ref{correlation_length}(b). This dynamical scaling indicates the emergence of a NTFP.

The universality class of the NTFP can be classified by the exponents $\beta$ and $\gamma$ in Eq.~\eqref{NTFP}, which are derived from power laws $L_{\rm c}(t) \propto t^{\beta}$ and $C_{\rm N}(0,t) \propto t^{\gamma}$ \cite{scaling}. Figures~\ref{correlation_length}(c) and (d) show $L_{\rm c}(t)$ and $C_{\rm N}(0,t)$ for four sets of parameters $U_{\rm 2}$ and $\Gamma$, from which we obtain $\beta \simeq 0.32$ and $\gamma \simeq 0.11$ by the least-square fit. These exponents are independent of the parameters and thus universal.

These exponents have never been reported in literature. To compare them with the wavenumber representation of previous studies \cite{NTF_exp1, NTF_exp2,NTF_GP1,NTF_GP2,NTF_GP3,NTF_GP4,NTF_GP5,NTF_GP6,NTF_GP7}, we perform the Fourier transformation of Eq.~\eqref{NTFP} in 1D systems, obtaining $ \bar{C}(k,t)=t^{\alpha} g(k t^{\beta}) $ with some function $g(x)$. Then, we find an exponent $\alpha=\gamma + \beta$ with $\alpha \simeq 0.43$, which does not agree with the results in Refs.~\cite{NTF_exp1, NTF_exp2,NTF_GP1,NTF_GP2,NTF_GP3,NTF_GP4,NTF_GP5,NTF_GP6,NTF_GP7}. 

\begin{figure}[b]
\begin{center}
\includegraphics[keepaspectratio, width=8.8cm,clip]{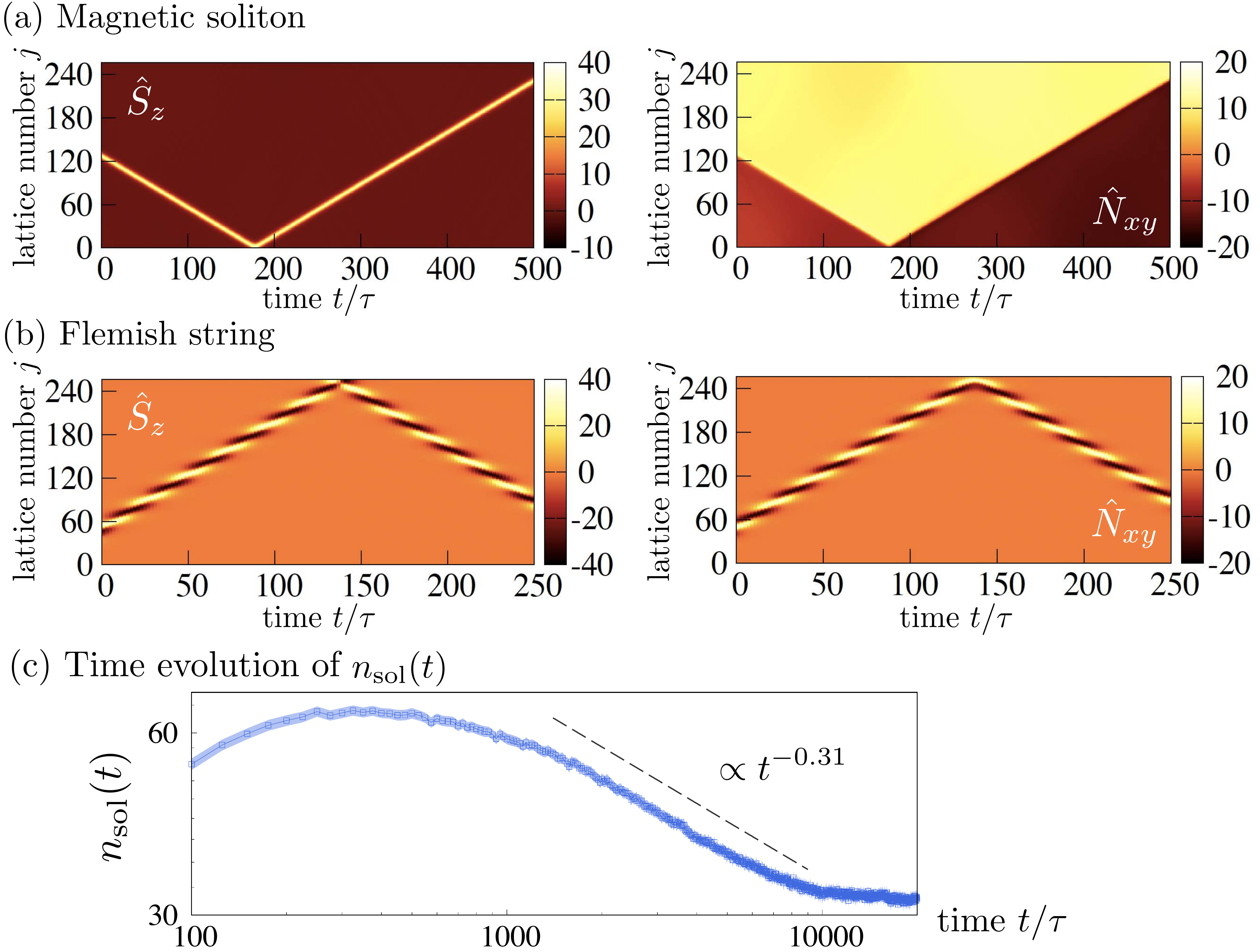}
\caption{(a) Magnetic soliton and (b) Flemish string. The left and right panels show the spatio-temporal distributions of $\hat{S}_{z,j}$ and $\hat{N}_{xy,j}$ obtained from the mean-field solution to Eq.~\eqref{spinor_BH} subject to the Neumann boundary condition. (a) A magnetic soliton is a locally magnetized stable object and separates the nematic order into two regions. (b) A Flemish string is a bound state of two magnetic solitons which periodically change their relative positions. In contrast to (a), it does not divide the nematic order. (c) Time evolution of a measure of the magnetic-soliton number $n_{\rm sol}(t)$ for $U_{2}/U_{0}=0.05$ and $\Gamma=2$, showing $n_{\rm sol} \propto t^{-0.31}$, which is almost the inverse of $L_{\rm c}(t) \propto t^{\beta}$ found in Fig.~\ref{correlation_length}(d). Color bands show $3\sigma$ error bars in the TWA calculations.  \label{solitons} }
\end{center}
\end{figure}

The universality of the thermalization dynamics emerges from two different types of soliton solutions, i.e., magnetic solitons and Flemish strings. The magnetic soliton is a solution of the integrable Gardner equation \cite{Gardner1}, which can be derived from application of singular perturbation theory to the continuous classical model of Eq.~\eqref{spinor_BH} \cite{magsoli1} (see Supplemental Material \cite{supl}). As shown in Fig.~\ref{solitons}(a), it has a locally magnetized stable object and separates the two different nematic orders, thereby decreasing the nematic correlation length. The thermalization dynamics would be promoted if a single magnetic soliton could be annihilated. However, it is, in fact, stable and cannot disappear spontaneously.

The annihilation of solitons proceeds through that of a Flemish string. This is a soliton solution of the integrable Manakov equation \cite{Manakov1,Manakov2,Manakov3,Manakov4}, which is a continuous classical model of Eq.~\eqref{spinor_BH} with $U_2=0$ (see Supplemental Material \cite{supl}). In contrast to magnetic solitons, it does not separate the nematic order as shown in Fig.~\ref{solitons}(b). This string is unstable for $U_{2} \neq 0$, so that it eventually vanishes and nematic domains can grow. We have confirmed that Flemish strings are formed through collisions of magnetic solitons, as indicated by the solid curves in Fig.~\ref{distribution}(b). The formation of Flemish strings and their subsequent decay constitute the main mechanism for soliton annihilation. The details including the lifetime and the number of Flemish strings are discussed in Supplemental Material \cite{supl}.

Another numerical evidence for the nematic domain growth due to solitons is the number of magnetic solitons. We calculate $n_{\rm sol} \coloneqq \sum_{j=0}^{M-1} \langle \theta(\hat{S}_{z,j}^2-S_{\rm th}^2) \rangle$ with $S_{\rm th}=0.7\rho_{\rm b}$ for the bulk particle number $\rho_{\rm b}$ \cite{soliton_number}, where $\theta(x)$ is the unit-step function. This should be proportional to $t^{-\beta}$ if $L_{\rm c}(t)$ obeys  the power law $t^{\beta}$. We find $n_{\rm sol} \propto t^{-0.31}$ as shown in Fig.~\ref{solitons}(c), which is consistent with $\beta \simeq 0.32$ obtained from $L_{\rm c}(t)$. This confirms the fact that annihilation of magnetic solitons is essential in the universal thermalization dynamics through the NTFP. This universal behavior appears when many solitons are generated. However, for longer times, the universal relaxation stops and $L_{\rm c}(t)$ and $C_{\rm N}(0,t)$ saturate because the number of solitons becomes small. 

\begin{figure}[t]
\begin{center}
\includegraphics[keepaspectratio, width=8.8cm,clip]{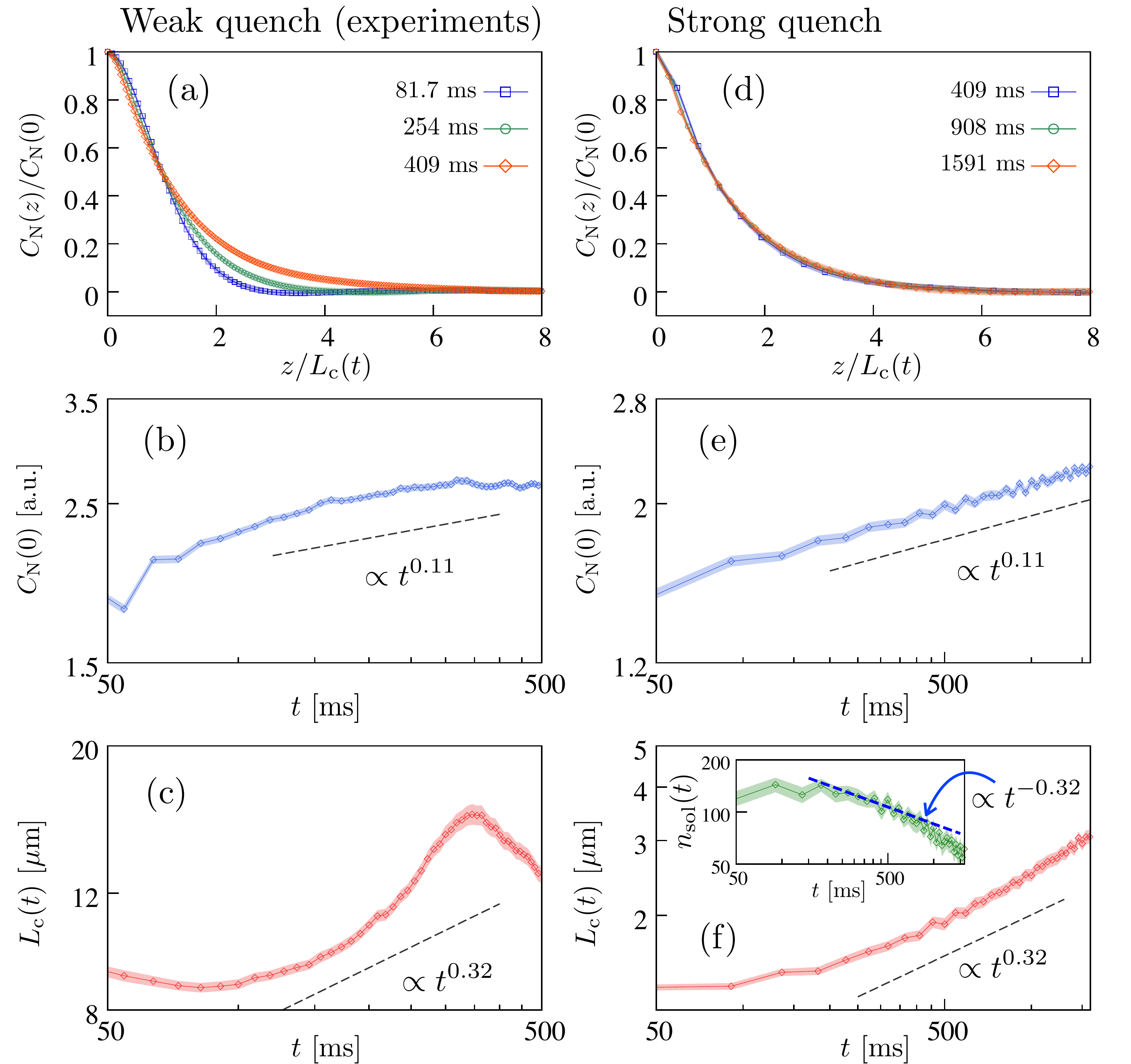}
\caption{Numerical results for (a) $C_{\rm N}(z,t)$, (b) $C_{\rm N}(0,t)$, and (c) $L_{\rm c}(t)$ following weak quench with the experimental parameters of Refs.~\cite{AF_quench1,AF_quench2}. They show neither the universal thermalization dynamics characterized by the dynamical scaling for $C_{\rm N}(z,t)$ nor the power law for $L_{\rm c}(t)$. (e)-(f) show the corresponding results for strong quench. We confirm the dynamical scaling and the power laws with the exponents found in the uniform system (See Fig.~\ref{correlation_length}). The inset in (f) shows $n_{\rm sol} \propto t^{-0.32} \propto  L_{\rm c}(t)^{-1}$ which is a measure of the number of magnetic solitons. Color bands show $3\sigma$ error bars in the TWA calculations.
\label{exp1}}
\end{center}
\end{figure}

{\it Analysis~of~the~experimental~thermalization~dynamics.-}
Let us examine the experiments \cite{AF_quench1,AF_quench2} where the thermalization dynamics in trapped 1D AF spinor Bose gases was investigated using density correlation functions. One is naturally led to ask whether or not the observed thermalization dynamics is universal. By numerically solving the spin-1 Gross-Pitaevskii (GP) equation with initial noises in the finite-temperature TWA method \cite{TWA1,FT_TWA1,FT_TWA2,FT_TWA3}, we obtain results in good agreement with the experiments \cite{supl}. We calculate the correlation function $C_{\rm N}(z,t)$ with the parameters used in the experiments as shown in Fig.~\ref{exp1}(a) \cite{definition_of_N}. Although $C_{\rm N}(0,t)$ shows the power-law behavior in a short time ($200 \sim 350~{\rm ms}$), we find neither the dynamical scaling of $C_{\rm N}(z,t)$ nor the power law of $L_{\rm c}(t)$. We therefore conclude that the universal thermalization dynamics does not emerge in the experiments.  

The absence of the universal scaling in Eq.~\eqref{NTFP} is attributed to a weak quench in the experiments, where $q(t)$ is quenched from $2.8~{\rm Hz}$ to $-4.2~{\rm Hz}$ \cite{exp_comment}. The energy scale of this change is given in terms of the spin interaction coefficient $c_{\rm 2}'$ of the GP model and the bulk density $\rho_{\rm b,exp}$ by $0.06 c_{\rm 2}' \rho_{\rm b,exp}$, which is too small to excite a sufficient number of magnetic solitons. 

To investigate the universal thermalization dynamics, we need a strong quench whose energy scale is of the order of the spin interaction energy $c_{\rm 2}' \rho_{\rm b,exp}$. We numerically quench $q$ from $2.8~{\rm Hz}$ to $-117~{\rm Hz}$. Figure~\ref{exp1}(b) shows the dynamical scaling in Eq.~\eqref{NTFP} and the power laws with the same exponents as those of the uniform system. Hence, we can confirm the universal thermalization dynamics through the NTFP. These results are consistent with the NTFP scenario in Fig.~\ref{intro}(a), where the strong quench leads to the NTFP but the weak one does not. While the spatial distribution of $\hat{N}_{xy,j}$ has not been observed, $\hat{S}_{z,j}$ can be measured experimentally. Thus, it is possible to estimate the exponent $\beta$ from $n_{\rm sol}$. The inset of Fig.~\ref{exp1}(b-3) confirms $n_{\rm sol} \propto t^{-\beta}$, which is an experimentally observable signature of the NTFP. 
 
$Conclusion.-$
By numerically studying the 1D thermalization in an antiferromagnetic spin-1 Bose gas, we find the universal nematic thermalization dynamics through a NTFP with scaling exponents $\beta \simeq 0.32$ and $\gamma \simeq 0.11$ in Eq.~\eqref{NTFP}. We identify the thermalization mechanism to be annihilation of Flemish strings of magnetic soliton pairs. The universal thermalization dynamics is discussed in the experimental settings in Refs.~\cite{AF_quench1,AF_quench2}. In 1D ultracold atomic gases, several different types of solitons other than magnetic solitons have experimentally been created \cite{soliton_exp1,soliton_exp2,soliton_exp3,soliton_exp4,soliton_exp5,soliton_exp6,soliton_exp7,soliton_exp8,soliton_exp9} and their dynamics has extensively been investigated \cite{atomic_gas_soliton,soliton1,soliton2,soliton3,soliton4,soliton5,soliton6,soliton7,soliton8,soliton9,soliton10,soliton11,soliton12,soliton13,soliton14,soliton15,soliton16}. It is interesting to investigate other NTFP universality classes with these solitons. 

\begin{acknowledgments}
We thank C. Raman for providing us the experimental data and useful comments on the manuscript, and I. Danshita, S. Furukawa, and M. Kunimi for fruitful discussions. This work was supported by KAKENHI Grant No. JP18H01145 and a Grant-in-Aid for Scientific Research on Innovative Areas ``Topological Materials Science'' (KAKENHI Grant No. JP15H05855) from the Japan Society for the Promotion of Science. R. H. was supported by the Japan Society for the Promotion of Science through Program for Leading Graduate Schools (ALPS) and JSPS fellowship (JSPS KAKENHI Grant No. JP17J03189). K. F. was supported by JSPS fellowship (JSPS KAKENHI Grant No. JP16J01683).
\end{acknowledgments}

\widetext
\clearpage

\setcounter{equation}{0}
\setcounter{figure}{0}
\setcounter{section}{0}
\renewcommand{\theequation}{S-\arabic{equation}}
\renewcommand{\thefigure}{S-\arabic{figure}}

\section*{Supplemental Material for ``Flemish Strings of Magnetic Solitons and a Non-Thermal Fixed Point in a One-Dimensional Antiferromagnetic Spin-1 Bose Gas''}

We describe integrable models for magnetic solitons and Flemish strings, discuss the main mechanism for magnetic-soliton annihilation, and provide details of our numerical analysis of the experiments of Refs.~\cite{AF_quench_S1,AF_quench_S2}.

\section{Integrable models for magnetic solitons and Flemish strings}

\subsection{Continuum approximation}
A classical trajectory in the truncated Wigner approximation (TWA) obeys the following classical equation of motion derived from Eq.~(2) in the main text:
\begin{eqnarray}
i \hbar \frac{\partial }{\partial t} b_{m,j} = &-&J (b_{m,j+1} + b_{m,j-1}) - q m^2 b_{m,j} - (2U_0 - U_2)b_{m,j}  \nonumber \\ 
&+& U_{0}  \rho_{m,j} b_{m,j} + U_{2}\sum_{n=-1}^1  \bm{S}_{j} \cdot (\bm{s})_{mn} b_{n,j}, 
\label{SpinorGP1}
\end{eqnarray}
where $b_{m,j}$, $\rho_j \coloneqq \sum_{m}  b_{m,j}^{*} b_{m,j}$  and $S_{\alpha,j} \coloneqq \sum_{m,n}  b_{m,j}^{*} ({s_{\alpha}})_{mn} b_{n,j}~(\alpha=x,y,z)$ with the spin-1 spin matrices $(s_{\alpha})_{mn}$ are c-numbers corresponding to the annihilation operators $\hat{b}_{m,j}$, the total particle number, and the spin vector, respectively. 

When the width of a soliton $l_{\rm s}$ is much larger than the lattice constant $a$, we can apply the continuum approximation to Eq.~\eqref{SpinorGP1}. In our numerical calculation for the thermalization dynamics in the main text, we have $l_{\rm s} \sim 9 a$. Denoting a macroscopic wave function by $\psi_m(x=ja) = b_{m,j}$, we derive the following spinor Gross-Pitaevskii (GP) equation: 
\begin{eqnarray}
i \frac{\partial }{\partial t} \psi_{m}(x,t) = -J'\frac{\partial^2}{\partial x^2} \psi_{m}(x,t) - q' m^2 \psi_{m}(x,t) + U_{0}'  \rho (x,t) \psi_{m}(x,t) + U_{2}'\sum_{n=-1}^1  \bm{S}(x,t) \cdot ({\bm{s}})_{mn}  \psi_{n}(x,t), \label{SpinorGP2}
\end{eqnarray}
where $J'=Ja^2/\hbar$, $q'=-q/\hbar$, $U_{0}'=U_{0}/\hbar$, $U_{2}'=U_{2}/\hbar$, $\rho (x,t) = \sum_{m=-1}^1 |\psi_{m}(x,t)|^2$, and $\bm{S} (x,t) = \sum_{n,m=-1}^1 \psi_{n}(x,t)^* ({\bm{s}})_{nm} \psi_{m}(x,t)$. Here, we eliminate the chemical potential term proportional to $b_{m,j}$ by applying the global gauge transformation: $\psi(x,t) \rightarrow \psi(x,t) {\rm exp}(-i W t)$ with $W=2J+2U_0-U_2$. 

In the quench dynamics considered in the main text, the parameter $q$ is quenched from the polar phase to the antiferromagnetic phase, so that the particle number of the $m=0$ component is very small. Therefore, we can neglect $\psi_0(x,t)$ in Eq.~\eqref{SpinorGP2}, obtaining
\begin{eqnarray}
i \frac{\partial }{\partial t} \psi_{1}(x,t) = &-& J'\frac{\partial^2}{\partial x^2} \psi_{1}(x,t) - q' \psi_{1}(x,t) \nonumber  \\ 
&+& U_{0}'  \big( |\psi_{1}(x,t)|^2 + |\psi_{-1}(x,t)|^2 \big)  \psi_{1}(x,t) + U_{2}'  \big( |\psi_{1}(x,t)|^2 - |\psi_{-1}(x,t)|^2 \big)  \psi_{1}(x,t), 
\label{SpinorGP3}
\end{eqnarray}
\begin{eqnarray}
i \frac{\partial }{\partial t} \psi_{-1}(x,t) = &-& J'\frac{\partial^2}{\partial x^2} \psi_{-1}(x,t) - q' \psi_{-1}(x,t) \nonumber  \\ 
&+& U_{0}'  \big( |\psi_{1}(x,t)|^2 + |\psi_{-1}(x,t)|^2 \big)  \psi_{-1}(x,t) - U_{2}'  \big( |\psi_{1}(x,t)|^2 - |\psi_{-1}(x,t)|^2 \big)  \psi_{-1}(x,t).
\label{SpinorGP4}
\end{eqnarray}
These coupled equations have some soliton solutions as explained in the following. 

\subsection{The Gardner equation}
Let us assume that the system has small but finite fluctuations for the amplitude and phase of $\psi_{m}(x,t)$. Then, we can derive an integrable equation called the Gardner equation using a singular perturbation method. We follow the derivation of Ref.~\cite{magsoli_S1}. Firstly, we write the wavefunction as
\begin{eqnarray}
    \begin{pmatrix}
       \psi_{1} (x,t)\\
       \psi_{-1} (x,t)
    \end{pmatrix}
  = \sqrt{\rho_{\rm tot}(x,t)} e^{i \Phi(x,t)/2}
    \begin{pmatrix}
      {\rm cos} (\theta (x,t)/2) e^{-i \phi (x,t)/2} \\
      {\rm sin}  (\theta (x,t)/2) e^{ i \phi (x,t)/2}
    \end{pmatrix}
\end{eqnarray}
with the local total particle number $\rho_{\rm tot} = |\psi_{1}|^2 + |\psi_{-1}|^2$, the angle $\theta$ determining the relative particle number, the global phase $\Phi$, and the relative phase $\phi$. Secondly, by employing the singular perturbation method under the assumption that fluctuations of $\rho_{\rm tot}$, $\theta$, $\Phi$, and $\phi$ are small but nonzero, we obtain an effective equation of motion of $\theta' = \theta -\pi/2$ given by
\begin{eqnarray}
\frac{\partial}{\partial t} \theta' + A \frac{\partial }{\partial x} \theta'+ B  (\theta')^2  \frac{\partial }{\partial x} \theta'  + C \frac{\partial^3 }{\partial x^3} \theta' = 0, 
\end{eqnarray}
where
\begin{eqnarray}
A =  \sqrt{ 2J' \bar{\rho} U_{2}'}, 
\end{eqnarray}
\begin{eqnarray}
B =  -\frac{3A(8U_0'+10U_2')}{8(U_0'-U_2')},
\end{eqnarray}
\begin{eqnarray}
C = -\frac{J'^3}{2A}, 
\end{eqnarray}
and $\bar{\rho}$ is the average particle number. This equation is a special case of the integrable Gardner equation. It has a magnetic-soliton solution, which is called a polarization Gardner soliton in Ref.~\cite{magsoli_S1}. We note that the magnetic-soliton solution can be analytically derived without resort to the singular perturbation method if the local total particle number $\rho_{\rm tot}$ is independent of space \cite{magsoli_S2}.  

\subsection{The Manakov equation}
When we choose $U_2'=0$, Eqs.~\eqref{SpinorGP3} and \eqref{SpinorGP4} become
\begin{eqnarray}
i \frac{\partial }{\partial t} \psi_{1}(x,t) = &-& J'\frac{\partial^2}{\partial x^2} \psi_{1}(x,t) - q' \psi_{1}(x,t) + U_{0}'  \big( |\psi_{1}(x,t)|^2 + |\psi_{-1}(x,t)|^2 \big)  \psi_{1}(x,t), 
\label{SpinorGP5}
\end{eqnarray}
\begin{eqnarray}
i \frac{\partial }{\partial t} \psi_{-1}(x,t) = &-& J'\frac{\partial^2}{\partial x^2} \psi_{-1}(x,t) - q' \psi_{-1}(x,t) + U_{0}'  \big( |\psi_{1}(x,t)|^2 + |\psi_{-1}(x,t)|^2 \big)  \psi_{-1}(x,t).
\label{SpinorGP6}
\end{eqnarray}
These coupled equations are integrable and known as the Manakov equation \cite{Manakov_S1}. This equation has soliton solutions, one of which is a Flemish string. The mathematical expression of the solution is given in Ref.~\cite{Manakov_S2}

\section{Mechanism for magnetic-soliton annihilation}

\begin{figure}[t]
\begin{center}
\includegraphics[keepaspectratio, width=17.5cm,clip]{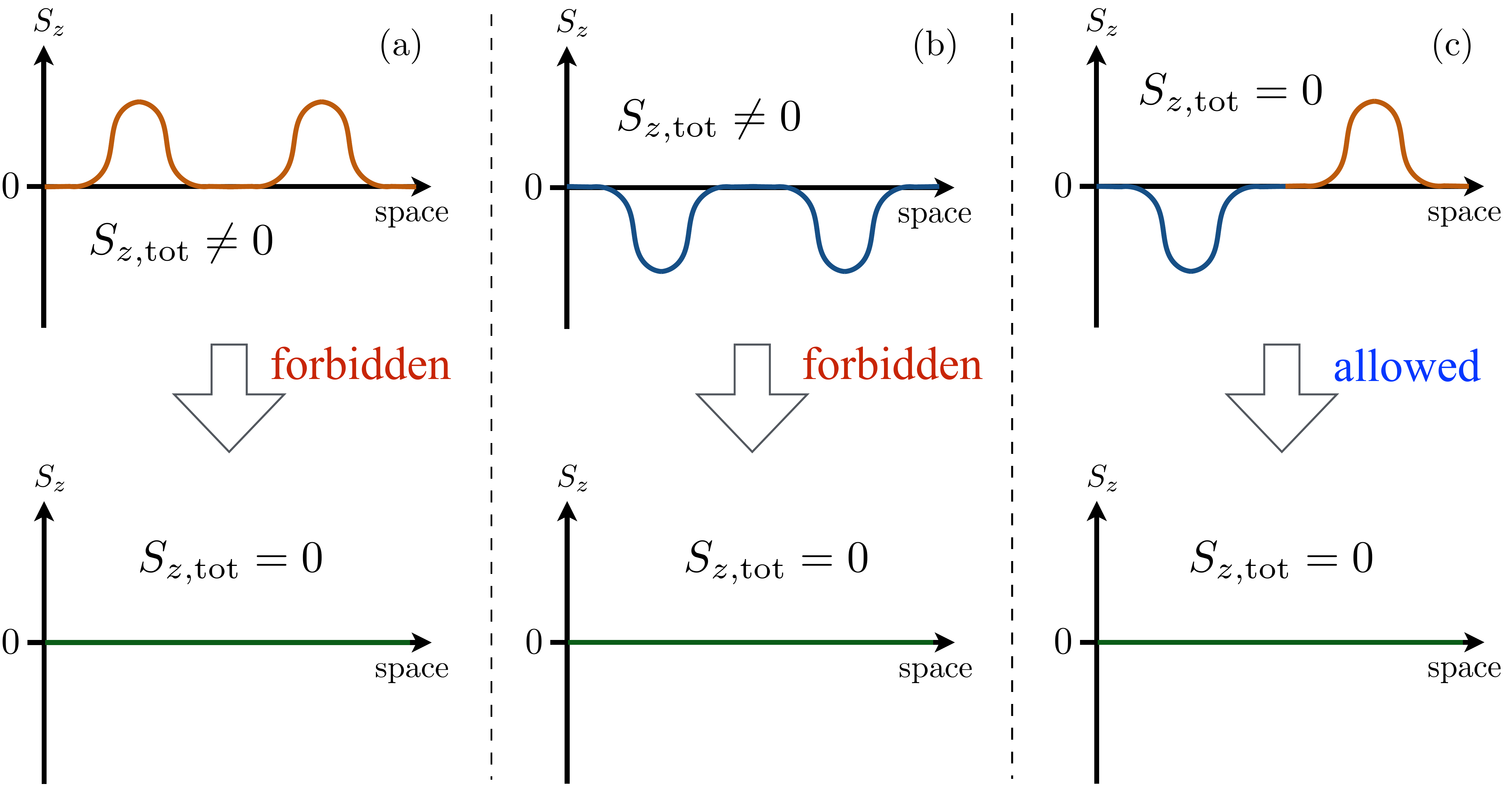}
\caption{Schematic illustrations for two magnetic-soliton collisions and the spin conversation of $S_{z,{\rm tot}}$. We consider three types of collision processes between two magnetic solitons with (a) only positive $S_z$, (b) only negative $S_z$, and (c) positive and negative $S_z$. The spin conversation prohibits the annihilation processes in (a) and (b), but does not in (c). This implies that the main mechanism for soliton annihilation should be the case of (c).
\label{s_fig0_1} }
\end{center}
\end{figure}

We discuss the main mechanism for magnetic-soliton annihilation in our universal relaxation. First, we point out the crucial role played by spin conservation and integrability breaking in the spin-1 Bose-Hubbard model, and then explain why formation of Flemish strings through magnetic-soliton collisions and their subsequent decay constitute the main mechanism for magnetic-soliton annihilation. Second, we numerically confirm this annihilation process by counting the number of Flemish strings and nematic domains. 

Soliton annihilation requires collision processes due to the stability of a single magnetic soliton. Our numerical calculations show that a single magnetic soliton is quite stable, and that any spontaneous single-soliton annihilation occurs. If we strongly excite the system with energy injection much larger than the spin exchange interaction, magnetic solitons might be broken into large amplitude spin waves. However, we do not inject such high energy into the system in the relaxation dynamics considered here, so that the spontaneous soliton annihilation should not occur. Thus, collision processes are needed for soliton annihilation.

First, let us make a physical argument about the annihilation process in light of spin conservation. Our system respects spin rotational symmetry about the z-axis, so that the spatially averaged magnetization $S_{z,{\rm tot}}$, defined in the following, is conserved:
\begin{eqnarray}
S_{z,{\rm tot}} = \frac{1}{M}\sum_{j=0}^{M-1} S_{z,j}.
\end{eqnarray}
Thus, collision processes of two magnetic solitons with the same sign of $S_z$ (see Figs.~\ref{s_fig0_1} (a) and (b)) are generally forbidden. As shown below, these processes are indeed not found in our numerical simulations. On the other hand, a collision process with opposite signs of $S_z$ as in Fig.~\ref{s_fig0_1}(c) can lead to annihilation of two magnetic solitons without breaking spin conservation. Therefore, spin conservation imposes a strong constraint on the soliton-annihilation mechanism.

\begin{figure}[t]
\begin{center}
\includegraphics[keepaspectratio, width=15cm,clip]{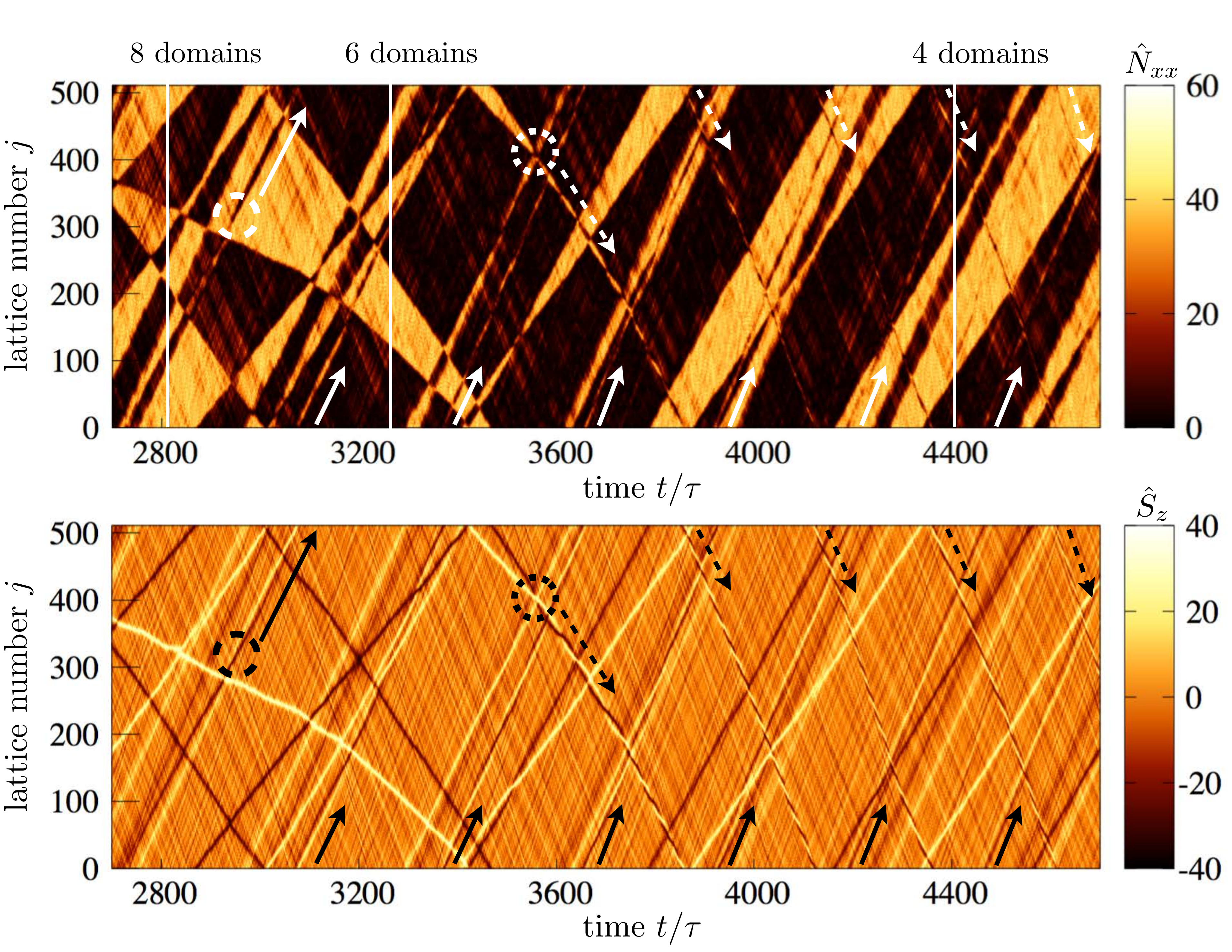}
\caption{Spatio-temporal distributions of nematicity (upper) and spin (lower). Two locations at which Flemish strings are formed are indicated by dashed circles, and the solid/dashed arrows show their directions of motion. Before the formation of Flemish strings, there are $8$ nematic domains. However, as magnetic solitons form Flemish strings, the number of nematic domains decreases. The first Flemish string appears around $t / \tau=2900$ as marked by the while dashed circle. We count the number of domains around $t/ \tau=3200, 4400$ as shown at the top of the figure. As time goes by, the amplitudes of the Flemish strings become smaller and gradually disappear. Everytime a Flemish string disappear, a magnetic soliton and an antisoliton annihilate pairwise, as can be seen from a decrease in the number of nematic domains. \label{s_fig0_2} }
\end{center}
\end{figure}

It is notable that Flemish strings can be formed through the collision process in Fig.~\ref{s_fig0_1}(c) before soliton-pair annihilation. This is because our model is close to the integrable Manakov model as described in the main text and this model has an exact Flemish-string solution. However, our model is not exactly integrable due to the weak spin-exchange interaction, and the Flemish string can disappear with a finite lifetime. Thus, this integrability breaking eventually leads to the soliton-pair annihilation. From this discussion, the time scale for the instability of Flemish strings is roughly estimated to be $\hbar/U_2$ with the spin-exchange interaction energy $U_2$. The quantitative lifetime might be analytically obtained by implementing soliton perturbation calculations \cite{soliton_cal}, from which we may derive an analytical expression of the lifetime. However, the investigation of this problem is far beyond the scope of the present work and we would like to leave it as a future problem. 

Summarizing the above discussion, we conclude that the main mechanism for the soliton annihilation is the collision process between the positive and negative magnetic-solitons that form unstable Flemish strings and the subsequent decay. In this mechanism, the spin conservation and the integrability breaking for the integrable Manakov model play essential roles.

We perform numerical simulations to investigate the annihilation processes of magnetic solitons, and find that two magnetic solitons with opposite signs disappear only through formation of a Flemish string. In the numerical calculation, preparing an initial state with 8 magnetic solitons (8 nematic domains), we solve the classical equation corresponding to the spin-1 Bose-Hubbard model with $U_0/J=0.025$ and $U_2/J=0.0025$ under a periodic boundary condition. The time is normalized by the characteristic time $\tau = 4\hbar /J$. Figure~\ref{s_fig0_2} shows spatio-temporal distributions of nematicity and spin, where there are initially 8 nematic domains divided by magnetic solitons. As time goes by, Flemish strings, indicated by arrows, are formed. Counting the number of nematic domains under the periodic boundary condition, one can confirm that the number decreases via formation of Flemish strings. For example, the domain number is 6 around $t/ \tau= 3200$ after the first formation of a Flemish string ($t/ \tau = 2900$, solid arrows) and decreases to 4 around $t/ \tau= 4400$ after the second formation ($t/ \tau= 3500$, dashed arrows). The spin amplitudes of Flemish strings become smaller in time and gradually disappear. Here, the time scale in which the amplitude starts to change is of the order of $\hbar/U_2 \sim 100 \tau$, which is consistent with our discussion of the Flemish-string instability based on the integrability breaking caused by the weak spin-exchange interaction. Our numerical calculations show no evidence of other annihilation mechanisms for magnetic solitons.

Finally, we discuss the number of Flemish strings. Although it is difficult to automatically identify the number by a computer, we can count the number by eye and confirm that the nematic order indeed becomes larger as the Flemish strings are formed and subsequently decay as shown in Fig.~\ref{s_fig0_2}. For example, one can find that two Flemish strings are formed around $t/ \tau=2900, 3500$, and that their subsequent decay leads to a decrease in the number of solitons. This is a clear evidence for the growth of nematic order through formation of Flemish strings. The reason for the difficulty of the automatic identification by a computer is that we need to scrutinize the time evolution of magnetic solitons to judge whether they are bound or not but we have yet to understand how to automatically judge twisting of Flemish strings.

\section{Numerical analysis for the experiments}

\subsection{Model}
We consider a Bose-Einstein condensate of $^{23}{\rm Na}$ with mass $M$ in a cigar-shaped harmonic trapping potential. To investigate the quench dynamics in this system, we use the GP equation given by
\begin{eqnarray}
i \hbar \frac{\partial}{\partial t} \psi_m (\bm{r},t) &=& \biggl( -\frac{\hbar^2}{2M} {\bm \nabla}^2 +  V_{\rm trap}(\bm{r}) + q(t) m^2 \biggl) \psi_m (\bm{r},t) \nonumber \\
&+& c_0 \rho({\bm r},t)  \psi_m (\bm{r},t) + c_2 \sum_{n=-1}^{1} {\bm S}({\bm r},t)\cdot (\bm{s})_{mn} \psi_n (\bm{r},t)
\label{3DGP}
\end{eqnarray}
with the macroscopic wavefunction $\psi_{m}(\bm{r},t)~(m=-1,0,1)$ and the quadratic Zeeman coefficient $q(t)$. The total particle density $\rho({\bm r},t)$, the spin density vector $\bm{S}({\bm r},t)$, and the trapping potential $V_{\rm trap}(\bm{r})$ are given by
\begin{eqnarray}
\rho({\bm r},t) = \sum_{m=-1}^{1} |\psi_m({\bm r},t)|^2, 
\end{eqnarray}
\begin{eqnarray}
\bm{S}({\bm r},t) = \sum_{m,n=-1}^{1} \psi_m({\bm r},t)^* (\bm{s})_{mn} \psi_n({\bm r},t), 
\end{eqnarray}
\begin{eqnarray}
V_{\rm trap}({\bm r}) = \frac{1}{2}M( \omega_z^2 z^2 + \omega_r^2 r^2 ), 
\end{eqnarray}
where $r=\sqrt{x^2+y^2}$ is the radial coordinate.
The parameters $c_0$ and $c_2$ are the strengths of spin-independent and spin-dependent interactions, which are expressed in terms of the s-wave scattering lengths $a_0$ and $a_2$ for the total-spin 0 and 2 channels by
\begin{eqnarray}
c_0 = \frac{4 \pi \hbar^2 (a_0 + 2 a_2)}{3M}, 
\end{eqnarray}
\begin{eqnarray}
c_2 = \frac{4 \pi \hbar^2 (a_2- a_0)}{3M}.
\end{eqnarray}

We adopt the following parameters used in the experiments of Refs~\cite{AF_quench_S1,AF_quench_S2}: $\omega_z=2 \pi \times 7~{\rm Hz}$, $\omega_r=2 \pi \times 470~{\rm Hz}$, and $M=23{\rm u}$, where $\rm u$ is the unified atomic mass unit. We choose $N_{\rm c}=9.4 \times 10^{6}$, $a_{0}=48.8a_{\rm B}$, and $a_{2}=51.2a_{\rm B}$ with the Bohr radius $a_{\rm B}$ such that the Thomas-Fermi radii $R_{r}$ and $R_{z}$ in the radial and axial directions, the bulk chemical potential $\mu_{\rm b}$, and the bulk spin interaction energy $\mu_{\rm s}$ agree with the experimentally observed values. Using these parameters, we obtain $R_{r} \sim 5.4~{\rm \mu m}$, $R_{z} \sim 360~{\rm \mu m}$, $\mu_{\rm b} \sim 354~{\rm nK}$, and $\mu_{\rm s} \sim h \times 120~{\rm Hz}$, where $h$ is the Planck constant. These values agree well with those observed experimentally \cite{AF_quench_S1,AF_quench_S2}. 

\subsection{Reduction of spatial dimensions to one}
Our system has a cigar-shaped configuration, so that we can reduce the three-dimensional system to a one-dimensional (1D) one. Because of tight confinement in the radial direction, we can express the wavefunction as
\begin{eqnarray}
\psi_m (\bm{r},t) = \phi_m(z,t) F(r), 
\label{3Dto1D}
\end{eqnarray}
where $\phi_m(z,t)$ is the one-dimensional wave function with magnetic quantum number $m$ and $F(r)$ is the radial wave function. As described in Ref.~\cite{AF_quench_S1}, when the energy scale of the quench is smaller than that of the first excited state in the radial direction, we can show that $F(r)$ is independent of time. 

Substituting Eq.~\eqref{3Dto1D} into Eq.~\eqref{3DGP}, we obtain the 1D GP equation given by
\begin{eqnarray}
i \hbar \frac{\partial}{\partial t} \phi_m (z,t) &=& \biggl( -\frac{\hbar^2}{2M} \frac{\partial^2}{\partial z^2} +   \frac{1}{2}M \omega_z^2 z^2 + qm^2 \biggl) \phi_m (z,t) \nonumber \\
&+& c_0' \rho_{\rm 1d}(z,t)  \phi_m (z,t) + c_2' \sum_{n=-1}^{1} {\bm S}_{\rm 1d}(z,t)\cdot (\bm{s})_{mn} \phi_n (z,t), 
\label{1DGP}
\end{eqnarray}
where $\rho_{\rm 1d}(z,t) = \sum_{m=-1}^{1} |\phi _m (z,t)|^2$ and $\bm{S}_{\rm 1d}(z,t) = \sum_{m,n=-1}^{1} \phi_m(z,t)^* (\bm{s})_{mn} \phi_n(z,t)$. The 1D interaction parameters are given by
\begin{eqnarray}
c_j' = c_j \Biggl(  \int dxdy |F(r)|^4 \Biggl)^{-1} ~~(j=0,2).
\end{eqnarray}

Finally, we consider the form of $F(r)$. The radial size of the condensate is $R_{r}\sim5.4~{\rm \mu m}$, which is much larger than the harmonic-oscillator length $a_{r}=\sqrt{\hbar/2M\omega_r} \sim 0.7~{\rm \mu m}$. Thus, we assume the Thomas-Fermi approximation for $F(r)$:
\begin{eqnarray}
F(r) =    \left\{
   \begin{aligned}
       \sqrt{ \frac{2}{\pi R_{\rm eff}^2} } \sqrt{1 - \frac{ r^2}{R_{\rm eff}^2 }} & ~~~~ (r<R_{\rm eff});  \\
       0 ~~~~ ({\rm otherwise})
   \end{aligned}
   \right.
\end{eqnarray}
with the effective radius $R_{\rm eff}$.
As a result, we obtain 
\begin{eqnarray}
c_j' = \frac{4c_j}{ 3 \pi R_{\rm eff}^2 } ~~(j=0,2).
\end{eqnarray}
Here, we use $R_{\rm eff} = 4.05~{\rm \mu m} \sim R_{r}$ which makes the physical quantities, such as the bulk chemical potential, agree well with those of the experiments in Refs.~\cite{AF_quench_S1,AF_quench_S2}. 

\subsection{How to generate initial states}
To solve Eq.~\eqref{1DGP}, we take the polar state as an initial state, which is generated by using the finite-temperature TWA method \cite{TWA0_S,TWA1_S,FT_TWA_S1,FT_TWA_S2,FT_TWA_S3}.

First, we numerically obtain a mean-field ground state by the imaginary-time evolution of Eq.~\eqref{1DGP}:
\begin{eqnarray}
\bm{\Phi}(z) = \left(
    \begin{array}{c}
      0 \\
      \Phi_0(z) \\
      0
    \end{array}
  \right), \label{initial1}
\end{eqnarray}
which satisfies
\begin{eqnarray}
-\frac{\hbar^2}{2M} \frac{\partial^2}{\partial z^2} \Phi_0(z) +   \frac{1}{2}M \omega_z^2 z^2 \Phi_0(z) + c_0' |\Phi_0(z)|^2  \Phi_0(z) = \mu \Phi_0(z), 
\end{eqnarray}
where $\mu$ is the chemical potential. Then, the initial state for Eq.~\eqref{1DGP} can be expressed as
\begin{eqnarray}
\phi_{1}(z) = \sum_{j=1}^{N_{{\rm cut},1}} \Bigl( R_{1,j} \alpha_{j}(z) + R_{-1,j}^*\beta_{j}(z)^*  \Bigl), 
\end{eqnarray}
\begin{eqnarray}
\phi_{0}(z) = \Phi_0(z) + \sum_{j=1}^{N_{{\rm cut},0}} \Bigl( R_{0,j}  u_{j}(z) + R_{0,j}^* v_{j}(z)^*  \Bigl), 
\end{eqnarray}
\begin{eqnarray}
\phi_{-1}(z) =  \sum_{j=1}^{N_{{\rm cut},-1}} \Bigl(  R_{-1,j} \alpha_{j}(z) + R_{1,j}^*\beta_{j}(z)^* \Bigl), 
\end{eqnarray}
where $N_{{\rm cut},m}$ is a cutoff to avoid the ultraviolet divergence, which is unavoidable when the TWA method is applied to continuous models \cite{TWA0_S,TWA1_S}. In our numerical calculation, the cutoff $N_{{\rm cut},m}$ is determined so that $E_{m,j} < \mu $ is satisfied. The functions $\alpha_{j}(z)$, $\beta_{j}(z)$, $u_{j}(z)$, and $v_{j}(z)$ obey the following Bogoliubov-de Gennes equations:
\begin{eqnarray}
  \begin{pmatrix}
  \mathcal{L}_1 & c_2' \phi(z)^2 \\
  -c_2' (\phi(z)^*)^2 & -\mathcal{L}_1
  \end{pmatrix}
  \begin{pmatrix}
  \alpha_j(z) \\
  \beta_j(z)
  \end{pmatrix}
  = E_{1,j}
  \begin{pmatrix}
  \alpha_j(z) \\
  \beta_j(z)
  \end{pmatrix},
\end{eqnarray}

\begin{eqnarray}
  \begin{pmatrix}
  \mathcal{L}_0 & c_0' \phi(z)^2 \\
  -c_0' (\phi(z)^*)^2 & -\mathcal{L}_0
  \end{pmatrix}
  \begin{pmatrix}
  u_j(z) \\
  v_j(z)
  \end{pmatrix}
  = E_{0,j}
  \begin{pmatrix}
  u_j(z) \\
  v_j(z)
  \end{pmatrix},
\end{eqnarray}

\begin{eqnarray}
 \mathcal{L}_1 = -\frac{\hbar^2}{2M} \frac{\partial^2}{\partial z^2} + \frac{1}{2} M \omega_z^2 z^2 + q - \mu + (c_0' + c_2') |\phi(z)|^2, 
\end{eqnarray}
\begin{eqnarray}
 \mathcal{L}_0 = -\frac{\hbar^2}{2M} \frac{\partial^2}{\partial z^2} + \frac{1}{2} M \omega_z^2 z^2 - \mu + 2c_0' |\phi(z)|^2.
\end{eqnarray}
The quantities $R_{1,j}$, $R_{0,j}$, and $R_{-1,j}$ are random numbers which are generated by the Wigner function defined by
\begin{eqnarray}
P(R_{m,j}, R_{m,j}^*) \propto {\rm exp} \biggl[ -2 |R_{m,j}|^2 {\rm tanh}\Bigl( E_{m,j}/k_{\rm B}T \Bigl) \biggl] ~~(m=-1,0,1)
\label{initial2}
\end{eqnarray}
with $E_{-1,j}=E_{1,j}$. 

\subsection{Comparison between experimental and numerical results}
We  perform numerical calculations of Eq.~\eqref{1DGP} with Eqs.~\eqref{initial1}-\eqref{initial2}, and compare the results with the experimental ones. Here, we compare the sum of the normalized particle numbers of the $m=1$ and $m=-1$ components and density fluctuations as a function of time. 
The former is defined by
\begin{eqnarray}
\mathcal{C}_{\rm sum}(t) = \frac{ N_1(t) + N_{-1}(t) }{N_{\rm tot}}, 
\end{eqnarray}
where $N_{\rm tot}$ is the total particle number and $N_m(t)$ is the particle number of the $m$-component. 
The latter are defined by
\begin{eqnarray}
\mathcal{D}_{m,n}(t) = \int dz \Bigl\langle  \Bigl[ {\rho}_{m}(z) - \langle {\rho}_{m}(z) \rangle \Bigl] \Bigl[ {\rho}_{n}(z) - \langle {\rho}_{n}(z) \rangle \Bigl] \Bigl\rangle, 
\end{eqnarray}
where ${\rho}_{m}(z) = |\phi_{m}(z)|^2$ is the particle density for the $m$-component, and the bracket denotes the ensemble average. 

\begin{figure}[t]
\begin{center}
\includegraphics[keepaspectratio, width=13cm,clip]{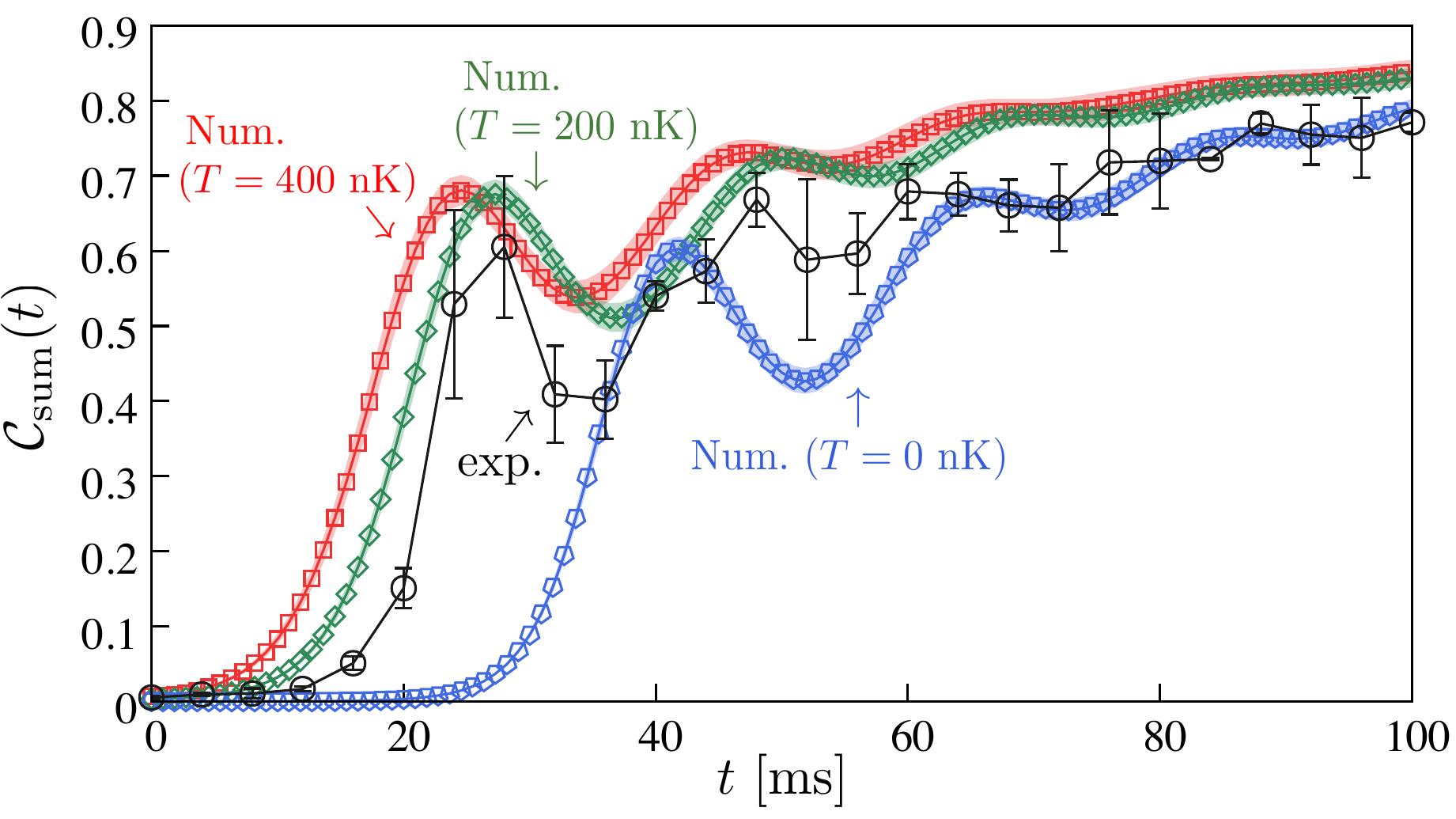}
\caption{Time evolution of $\mathcal{C}_{\rm sum}(t)$ at $T=0~{\rm nK}$ (blue), $200~{\rm nK}$ (green), and $400~{\rm nK}$ (red). The experimental results, taken from Fig.~1 in Ref.~\cite{AF_quench_S2}, are superimposed as the black circles with error bars. The temperature observed in the experiments is $400{\rm nK}$. In the early stage ($t<30 {\rm ms}$), the zero-temperature result largely deviates from the experiments, but the results at $T=200~{\rm nK}$ and $400~{\rm nK}$ agree better with it. In the later stage ($t>60~{\rm ms}$), all results qualitatively agree with the experimental data. Color bands show $3\sigma$ error bars in the ensemble average.  
\label{s_fig1} }
\end{center}
\end{figure}

Figure~\ref{s_fig1} shows the time evolution of $\mathcal{C}_{\rm sum}(t)$ and its dependence on temperature $T$. The temperature reported in the experiments of Refs.~\cite{AF_quench_S1,AF_quench_S2} is $T_{\rm exp} = 400~{\rm nK}$. We find that our numerical results at $200~{\rm nK}$ and $400~{\rm nK}$ show better agreement with the experimental data than that at the zero-temperature result in the early stage ($t<30~{\rm ms}$), but in the later stage ($t>60~$ms) the results at $T = 0~{\rm nK}$, $200~{\rm nK}$, and $400~{\rm nK}$ agree qualitatively with the data.  

\begin{figure}[t]
\begin{center}
\includegraphics[keepaspectratio, width=13cm,clip]{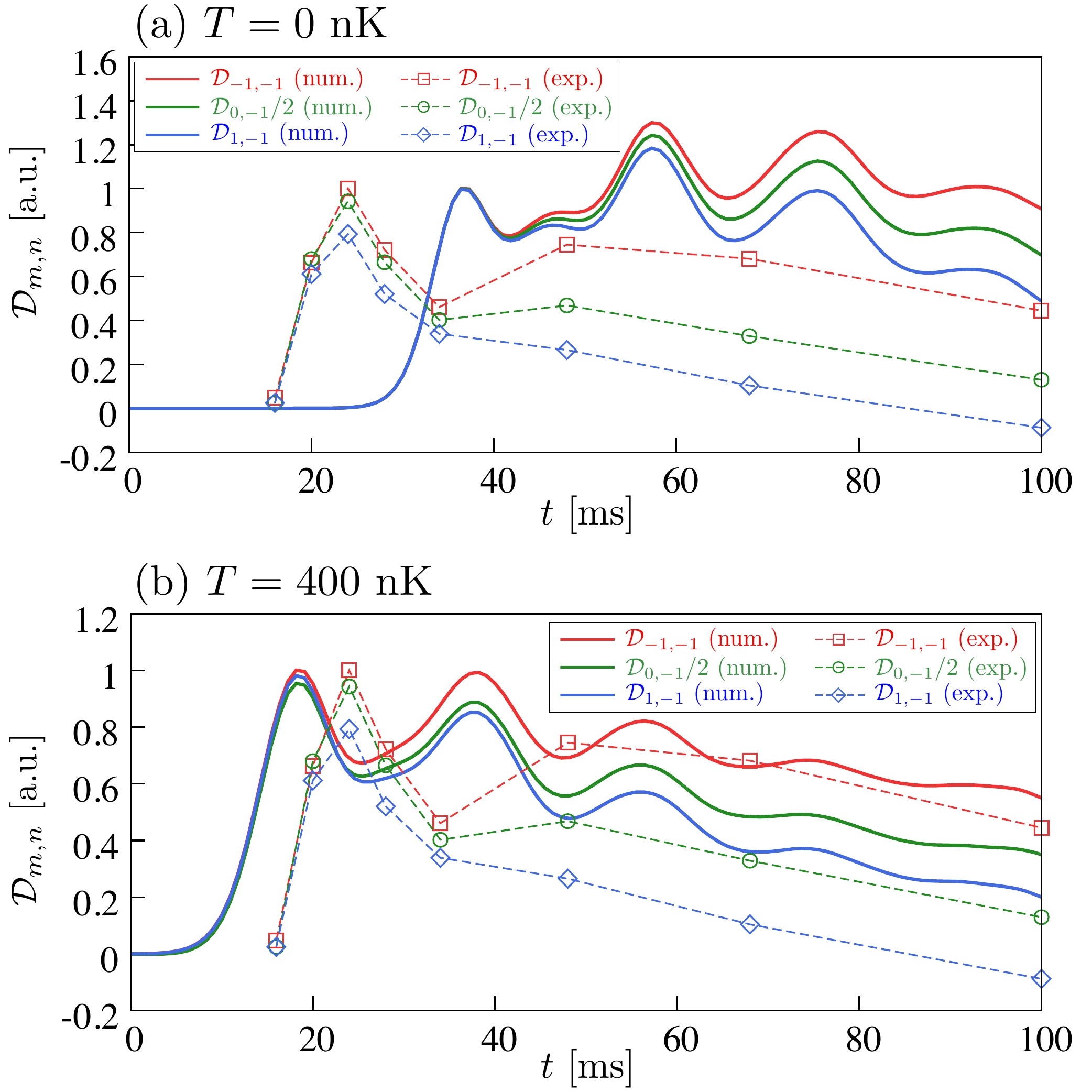}
\caption{Time evolution of the density fluctuations $\mathcal{D}_{m,n}(t)$ with red, green, and blue showing the data for $\mathcal{D}_{-1,-1}(t)$, $\mathcal{D}_{0,-1}(t)/2$, and $\mathcal{D}_{1,-1}(t)$, respectively. The three dashed lines show the experimental data, taken from Fig.~3 in Ref.~\cite{AF_quench_S1}, and the three solid curves show our numerical results. The results with $T=400~{\rm nK}$ show better agreement with the experiments. \label{s_fig2} }
\end{center}
\end{figure}

The numerical results of $\mathcal{D}_{m,n}(t)$ are shown in Fig.~\ref{s_fig2}, where we plot the two results at $T = 0~{\rm nK}$ and $400~{\rm nK}$. In contrast to Fig.~\ref{s_fig1}, the density fluctuations $\mathcal{D}_{m,n}(t)$ at = $400~{\rm nK}$ show the better agreement with the experimental results than those at zero temperature. Compared with the previous numerical results (Fig.~3 in Ref.~\cite{AF_quench_S1}), the agreement improves considerably. We suspect that the remaining deviation between our numerical results and the experimental ones may arise from the time-of-flight Stern-Gerlach observation in the experiments because it takes about $30~{\rm ms}$, which is comparable with the time scale needed before $\mathcal{C}_{\rm sum}(t)$ and $\mathcal{D}_{m,n}(t)$  in Figs.~\eqref{s_fig1} and \eqref{s_fig2} start to grow.

In summary, our numerical calculation (especially for $T=400~{\rm nK}$) can describe the experimental result well if we use the finite-temperature TWA method. Since the parameter region is in the deep superfluid regime where the method works well, we expect that our numerics can predict the thermalization dynamics observed in the experiments qualitatively beyond the experimental time scale $\sim100~{\rm ms}$ as well.

\subsection{Spatio-temporal distribution of $S_{z}$ and $N_{xy}$ in the quench dynamics}
We perform numerical calculations with two quench protocols: weak quench corresponding to the experiments of Refs.~\cite{AF_quench_S1,AF_quench_S2} and strong quench. We suddenly change $q$ from $2.8~{\rm Hz}$ to $-4.2~{\rm Hz}$ in the former case, and from $2.8~{\rm Hz}$ to $-117~{\rm Hz}$ in the latter case, and examine the spatio-temporal distributions of $S_{{\rm 1d},z}(z,t)$ and  $N_{{\rm 1d},xy}(z,t)$ for both cases, which are calculated from a single trajectory of Eq.~\eqref{1DGP}. Here, $N_{{\rm 1d},xy}(z,t)$ is given by 
\begin{eqnarray}
N_{ {\rm 1d},xy }(z,t) = \sum_{m,n=-1}^{1} \phi_m (z,t)^* (\hat{N}_{xy})_{mn} \phi_n (z,t),
\end{eqnarray}
\begin{eqnarray}
(\hat{N}_{xy})_{mn} = \frac{1}{2} \sum_{l}  \Bigl[(s_{\alpha})_{ml}(s_{\beta})_{ln}  + (s_{\beta})_{ml}(s_{\alpha})_{ln} \Bigl]. 
\end{eqnarray}

\begin{figure}[t]
\begin{center}
\includegraphics[keepaspectratio, width=17.5cm,clip]{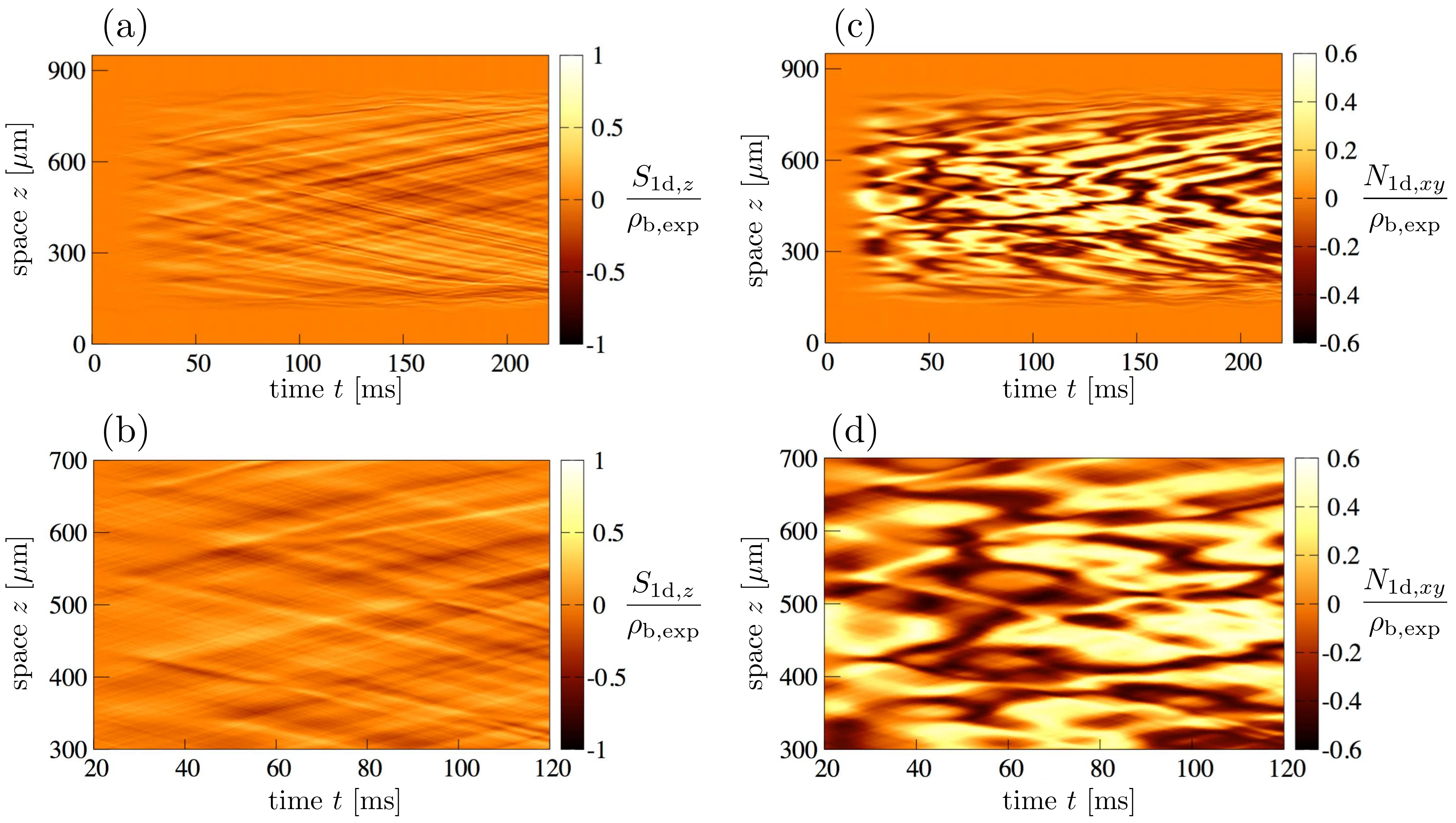}
\caption{Spatio-temporal distributions of (a,b) $S_{{\rm 1d},z}(z,t)$ and (c,d) $N_{{\rm 1d},xy}(z,t)$ for the experimental situation with (b) and (d) showing enlarged views corresponding to (a) and (c), respectively. Only a few magnetic solitons are excited. \label{s_fig3} }
\end{center}
\end{figure}

\begin{figure}[t]
\begin{center}
\includegraphics[keepaspectratio, width=17.5cm,clip]{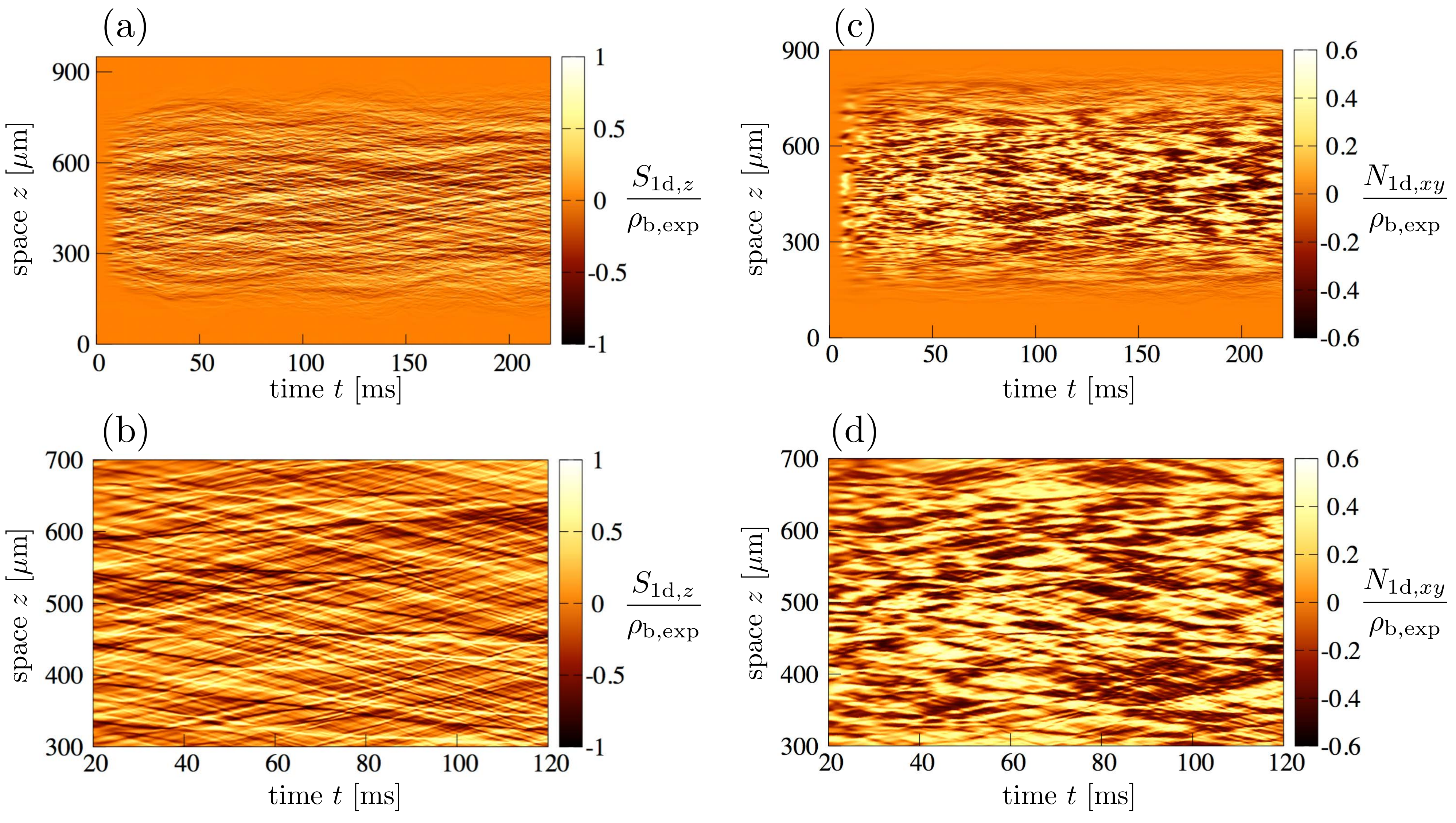}
\caption{Spatio-temporal distribution of (a,b) $S_{{\rm 1d},z}(z,t)$ and (c,d) $N_{{\rm 1d},xy}(z,t)$ for the strong quench with (b) and (d) showing enlarged views corresponding to (a) and (c), respectively. Many solitons are excited compared with Fig. S-3. \label{s_fig4} }
\end{center}
\end{figure}

Figure~\ref{s_fig3} illustrates the spatio-temporal distributions for the weak quench. In this case, the energy scale of the quench $\sim 0.06c_2' \rho_{\rm b,exp}$ is much smaller than the spin-exchange interaction energy $c_2' \rho_{\rm b,exp}$ at the bulk density $\rho_{\rm b,exp}$, so that only a few magnetic solitons are generated. On the other hand, for strong quench, many magnetic solitons are created as shown in Fig.~\ref{s_fig4} because the energy scale for this quench is about $c_2' \rho_{\rm b,exp}$. 

\subsection{Discussion on the density correlation function observed in Ref.~\cite{AF_quench_S2}}
We discuss how the density correlation function observed in Ref.~\cite{AF_quench_S2} is related to our universal relaxation dynamics. The experiment is a sequel to Ref.~\cite{AF_quench_S1}, and studied the relaxation dynamics in an antiferromagnetic phase after quenching the quadratic Zeeman coupling $q$ from positive to negative values. By observing a spatial correlation function for a density distribution, the experiment confirmed that the correlation function exhibits relaxation to a stationary state.

We conclude that our universal relaxation and the power exponents are not directly related to the experimentally observed results because of the following reason. The experiment observed relaxation after the weak quench, and generated only a few magnetic solitons as shown in our numerical calculations in Fig.~\ref{s_fig3}. Thus, the observed relaxation is essentially different from our universal relaxation with many magnetic solitons, and there is no direct relation between our power exponents and the observed density correlation function. 

Finally, let us briefly discuss our system after a long time, although this is beyond the scope of our research. In this case, our system has only a few magnetic solitons, and becomes similar to the experimental situation. Thus, on this time scale, a relation might be established between the experiment and our system. Reference \cite{AF_quench_S2} has reported that the density correlation length decreases in time (the exponent $\beta$ is negative), which implies growth of fine density structures. By taking into account that only a few magnetic solitons are generated, this behavior reminds us of a cascade in wave turbulence \cite{WT1,WT2}, in which nonlinear interactions between waves with different wavenumbers generate finer structures. In the field of ultracold gases, several theoretical and experimental studies have found such turbulent cascades \cite{WT_BEC1,WT_BEC2,WT_BEC3,WT_BEC4,WT_BEC5,WT_BEC6,WT_BEC7,WT_BEC8,WT_BEC9}. We thus expect that a signature of a wave turbulent cascade may appear for few soliton cases in experiments or in the long-time regime of our setup. Applying weak wave turbulence theory \cite{WT1,WT2} to the spin-1 BH model, we could analytically derive several power exponents for density, spin, and nematic correlation functions, which might provide direct relation between the experimental and our theoretical results.

\thispagestyle{myheadings}


\begin{thebibliography}{99}
\bibitem{scaling} From Eq.~\eqref{NTFP} and definition of $L_{\rm c}(t)$, we obtain a relation $0.5 f(0) = f(L_{\rm c}(t) t^{-\beta}/a)$. As the left-hand side is independent of $t$, we obtain $L_{\rm c}(t) \propto t^{\beta}$.
\bibitem{soliton_number} The measure $n_{\rm sol}$ of the soliton number is a lattice number of regions where $\hat{S}_{z,j}^2$ is larger than $S_{\rm th}^2$. When we calculate the soliton number, $n_{\rm sol}$ should be divided by width of a soliton. The typical width is $3\sim8 $ lattice points.
\bibitem{definition_of_N} The correlation function $N_{xy}(z,t)$ is defined by replacing the lattice index $k$ and the summation $\sum _k$ in Eq.~\eqref{Ncorrelation_function} with the coordinate $z$ and the integral $\int dz$, respectively.
\end{thebibliography}
\end{document}